# Status and Performance of New Silicon Stripixel Detector for the PHENIX Experiment at RHIC: Beta Source, Cosmic-rays and Proton Beam at 120 GeV

**Rachid Nouicer**[a1], **Yasuyuki Akiba**[b,c], **Robert Bennett**[d], **Kieran Boyle**[b],
**Vince Cianciolo**[e], **Abhay Deshpande**[b,d], **Alan Dion**[f], **Michael Eggleston**[f],
**Akitomo Enokizono**[e], **Steve Kaneti**[d], **Eric J. Mannel**[g], **Craig Oglvie**[f], **Hua Pei**[f],
**Andrey Sukhanov**[a], **Swadhin Taneja**[d], **Manabu Togawa**[b,c], **and the PHENIX-VTX Collaboration**

[a] *Brookhaven National Laboratory, Physics Department, Upton New York, 11973-5000, U.S.A.*
[b] *RIKEN BNL Research Center, Brookhaven National Laboratory, Upton, NY 11973, U.S.A.*
[c] *RIKEN Nishina Center for Accelerator Based Science 2-1 Hirosawa, Wako, Saitama 351-0198, Japn*
[d] *Stony Brook University, Department of Physics and Astronomy, Stony Brook, NY 11794, U.S.A*
[e] *Oak Ridge National Laboratory, Oak Ridge, TN 37831, U.S.A.*
[f] *Iowa State University, Department of Physics and Astronomy, Ames, IA 56011, U.S.A.*
[g] *Columbia University, Nevis Laboratories, Irvington, New York, 10533, U.S.A.*

*E-mail*: rachid.nouicer@bnl.gov

**Abstract.** We are constructing a Silicon Vertex Tracker detector (VTX) for the PHENIX experiment at RHIC. Our main motivation is to enable measurements of heavy flavor production (charm and beauty) in *p+p, p+d* and *A+A* collisions. Such data will illuminate the properties of the matter created in high-energy heavy-ion collisions. The measurements also will reveal the distribution of gluons in protons from *p+p* collisions. The VTX detector consists of four layers of barrel detectors and covers $|\eta| < 1.2$, and almost a $2\pi$ in azimuth. The inner two silicon barrels consist of silicon pixel sensors; their technology accords with that of the ALICE1LHCB sensor-readout hybrid. The outer two barrels are silicon stripixel detectors with a new "spiral" design, and a single-sided sensor with 2-dimensional (X, U) readout. In this paper, we describe the silicon stripixel detector and discuss its performance, including its response to electrons from a beta source ($^{90}$Sr), muons from cosmic-rays, and a 120 GeV proton beam. The results from the proton beam demonstrate that the principle of two-dimensional position sensitivity based on charge sharing works; the signal-to-noise value is 10.4, the position resolution is 33.6 μm for X-stripixel (35.2 μm for U-stripixel), and the tracking efficiencies in the X- and U-stripixels are, over $98.9 \pm 0.2\%$. The stripixel detector within the VTX project is in the pre-production phase.

---

[1] Corresponding author. Tel.: +1 6312190468; fax: +1 6313441334

# Contents

## 1. Physics motivation

Research at the RHIC (Relativistic Heavy Ion Collider) at Brookhaven National Laboratory (BNL) has led to a major physics discovery, namely the creation in high-energy central gold-gold collisions of a new form of matter, the strongly coupled quark-gluon plasma, or sQGP [1]. This finding was rated the top physics story of 2005 and the four experiments at RHIC, BRAHMS, PHENIX, PHOBOS, and STAR published, in White Papers, evidence of the existence of this new form of matter [2]. The focus of research now has shifted from this initial discovery to a detailed exploration of partonic matter. Particles carrying a heavy flavor (charm or beauty quarks) like *D*- and *B*-mesons, represent powerful tools for studying the properties of the hot, dense medium created in high-energy nuclear collisions: generated early in the reaction, they subsequently diffuse through the created matter. Since the $D^{\pm}$ and $B^{\pm}$ mesons have a finite decay-length ($c\tau$), respectively of about 312 μm and 491 μm, the measurement heavy-flavor production begins identifying displaced vertices. These particles support the precise probing of the spin structure of the proton in a new way such as measuring gluon distribution *G(x)* over much wider x range [3]. RHIC-II, an upgrade of RHIC, will be completed by 2012, with an increase of average luminosity up to 40 x $10^{26}$ cm$^{-2}$ s$^{-1}$ for *Au+Au* collisions, and 2 x $10^{32}$cm$^{-2}$s$^{-1}$ for polarized proton-beams [4].

Impelled by the exciting physics we uncovered, and in parallel to the RHIC upgrade, the PHENIX collaboration [5] plans to improve its experiment with a detector to exploit opportunities to bring new physical phenomena within reach. Accordingly, we are constructing the Silicon Vertex Tracker (VTX) that will allow us to obtain measurements from detecting particles carrying a heavy flavor [3]. The VTX detector consists of four layers of barrel detectors located in the region of pseudo-rapidity |η| < 1.2, and covers almost a 2π azimuthal angle. Its acceptance extends over a much larger span than that of the PHENIX's central arms, |η|< 0.35 and Δϕ = π/2 (the central arms measure electrons). The inner two silicon barrels consist of silicon pixel sensors whose technology follows that of the ALICE1LHCb sensor-readout hybrid developed at CERN for the ALICE and the LHCb experiments [6]. The outer two barrels comprise a silicon stripixel detector with a new "spiral" design, and a single-sided sensor with a 2-dimensional (X-U) read-out. Ref. [3] summarizes the physical specifications of the VTX. The specifications were chosen such that the design has a low material budget to minimize multiple scattering, and photon conversions ($\gamma \rightarrow e^+ e^-$), two effects which complicate electron measurements in the outer detectors.

We first describe the elements of the silicon stripixel detector: the silicon sensor, and our prototype readout card. Then we discuss the response of the stripixel detector to a beta-source $^{90}$Sr (strontium), cosmic-rays, and a 120 GeV proton-beam. These findings quantify the signal-to-noise ratio, tracking efficiency (detection efficiency), and residual distribution (position resolution); we also give the status of the stripixel detectors.

## 2. Silicon stripixel detector

Fig. 1 shows the silicon stripixel detector (stripixel detector), consisting of one silicon sensor, one prototype readout-card (ROC), and twelve SVX4s readout chips [7]. The silicon sensor's analog signals are readout via SVX4 readout chips mounted on the ROC. The latter is connected to a prototype ROC control Chip (RCC), and a Front-End Module (FEM) compatible with the PHENIX DAQ system. The stripixel detector is mounted on a thermal plate of Carbon Fiber Composite (CFC) containing two cooling tubes; details are given below.

### 2.1 Silicon sensor "spiral stripixel sensor design"

Researchers at BNL have developed a novel stripixel silicon sensor, a single-sided, N-type, DC-coupled, two-dimensional sensitive detector [8]. This design simplifies fabrication for both the sensor and signal processing compared with that for the conventional double-sided strip sensor. Each pixel from the stripixel sensor consists of two implants (a-pixel and b-pixel separated by a 3 μm gap and both are p+ implants) interleaved such that both implants register the charge deposited by ionizing particles (Fig. 2a.) Each interleaved implant is 5 μm wide and makes 5 turns in each pixel. Fig. 2b shows the way that the a-pixels are connected to form a X-stripixel; Fig.2c depicts the connection between b-pixels that forms a U-stripixel. The stereoscopic angle between an X-stripixel and a U-stripixel is 4.6°. Due to stereoscopic readout, the effective pixel size is 0.08×1.0 (mm$^2$). The overall size of the silicon sensor is about 34.9± 0.04 mm wide, 63.6 ± 0.04 mm long, and 0.625 ± 0.015 mm thick. Each sensor has 1536 channels. The silicon sensors were mass-produced by Hamamatsu Photonics (HPK) company. Fig. 2d is a photograph of one sensor. Quality assurance tests were completed; the total leakage current per sensor is 179 nA, implying a leakage of 0.12 nA per stripixel which is very low and allows us to use SVX4 readout chips. The capacitance per stripixel is 1.5 pF. The full depletion voltage of the stripixel sensor is about 100 Volts, the operation voltage for all measurements was set to 200 Volts and the breakdown voltage is up to 350 Volts. Ref. [9] details the findings from these quality assurance tests.

### 2.2 Prototype readout-card (ROC)

The prototype readout card (ROC) is a thin printed circuit board 15.2 mm long, 65 mm wide, and 0.6 mm thick. It hosts twelve SVX4 readout chips that process the silicon sensor's charge signals, and a few passive components, such as a bias voltage filter capacitor, near each chip (Fig. 1). The backplane (ohmic side) of the silicon sensor is glued to the top of the ROC, which supplies the high voltage bias connection. The ROC was designed to minimize the impedance between the analog ground connections of different SVX4 chips, and to shield the analog ground plane from digital signals.

## 3. Performance of silicon stripixel detector

To ensure the long-term operation of the silicon stripixel detector, it is mandatory to study their performance. We assembled at BNL three prototype detectors using three mass-produced silicon sensors, three readout-cards (ROCs) and the SVX4 readout chips, and then wire bonded the chips to the ROCs. At Fermi National Accelerator Laboratory (FNAL) the SVX4 chips were wire bonded to the silicon sensor and the wires were encapsulated. Fig. 1 is a photograph of one of these three stripixel detectors. The detector is mounted on the Carbon Fiber Composite (CFC) thermal plate that has two cooling tubes and a hole in its middle (2.0 cm x 2.5 cm) that allows the passage of particles. The cooling fluid (70% water and 30% propylene glycol) is maintained at $0^o$ C.

We evaluated the performance of the stripixel detectors, including the whole readout system, using a beta source ($^{90}$Sr), a cosmic-ray and a proton beam at 120 GeV. Our main goal was to determine

- Signal-to-Noise ratio (S/N)
- Residual distribution (position resolution)
- Tracking efficiency (detection efficiency)

### 3.1 Response to electrons from a $^{90}$Sr beta source

Electrons from the $^{90}$Sr (strontium) beta source, generated from two different beta decays, have a wide energy spectrum with an end point at about 2.5 MeV. The results presented here were obtained with a simple setup. The detector was positioned between the beta source and the scintillators. Then the detector, including the thermal plate which was cooled to zero degrees, the beta source, and the scintillators were placed inside a dark box through which dry nitrogen flowed (the setup for the beta source test is similar to that used for cosmic-rays shown in Fig. 4). The acquisition system was triggered by a coincidence between the scintillators and the digital output of the stripixel detector, indicating that the silicon sensor was fired. In this preliminary test, the particle's energy was not selected.

Fig. 3 illustrates the response of stripixel detector to electrons from the $^{90}$Sr beta source; Fig. 3a gives the raw ADC values versus channel number. There is a clear pedestal distribution for each channel from the twelve SVX4 readout chips; the pedestal is mean value = 72 ADC counts and the pedestal's width (noise) ~ 9.6 ADC counts. We also observed that the electron signals from the beta source were located only in two regions (when the beta source was positioned under the silicon sensor), U2L and X2L, indicating a good correlation (a good charge-sharing). Fig. 3b plots the corrected ADC (energy loss) versus the channel number after applying an offline analysis method wherein the pedestal value was subtracted event-by-event, and the ADC value in each channel determined by averaging with the two neighboring channels' ADC (±2 channels). We plan to implement this method in the FPGA (the FPGA is located on the RCC board) and acquire data with zero suppression.

To find the charge-sharing peak in the X and U stripixels, we employed a clustering algorithm as follows: 1) To tag a channel as a hit, the corrected ADC value for that channel must exceed two sigma of the pedestal's width; 2) if the hit is shared with next channel, merging (summation) is applied so that the clustered ADC is the sum of both of them, and the clustered channel is calculated by the centroid method, and, 3) the clustered ADC must be greater than 40 ADC counts bias voltage filter capacitor to prohibit the inclusion of numbers of fake clusters formed by noise. The signal of the charge-sharing peaks in the X-stripixel and U-stripixel, respectively, are shown in Figs. 3c and d. The signal-to-noise (S/N) ratio in the X (or

U) stripixel is defined as the ratio of the mean value of the signal of the charge-sharing peak divided by the average width of the pedestal (noise). This gives the S/N ratio in the stripixel detectors using the beta-source, as

- X-stripixel channel: S/N = 9.7
- U-stripixel channel: S/N = 9.6

We note that the trigger was not well optimized for the wide energy spectrum of a beta source; particles from a beta source are emitted randomly and arise from two different beta decays. We expect slightly better results with a beam.

In conclusion, the response of the stripixel detector to electrons from beta source, $^{90}$Sr, is very encouraging; they uphold the basic principle of the functionality of this detector (charge sharing). and display good agreement between measured- and simulated-electron signal peaks.

### 3.2 Response to muons from cosmic-rays

Fig. 4 shows the setup of the cosmic-ray test. It is similar to that for the beta source; however, the dark box contains two silicon stripixel detectors positioned in parallel with three scintillators located outside the box. The triple coincidence signal generated by muons from the cosmic-ray crossing the three scintillators is the trigger for the readout system.

Fig. 5 shows a single deposition event of particle energy loss (corrected ADC) in the two stripixel detectors. The results depicted were obtained after applying the event-by-event pedestal subtraction method discussed in the section 3.1. For a single event, we observe clearly the muon track crossing the two stripixel detectors; there are two well-correlated hits (U1R and X1R) in stripixel detector-1, and two similar ones (U1L and X1L) in stripixel detector-2. The basic principles of tracking hits can easily be applied to the stripixel detectors.

### 3.3 Response to proton beam at 120 GeV

The response of the silicon stripixel detector to the proton beam at 120 GeV was carried-out at the Meson Test Beam Facility at FNAL (experiment T984) during August 2008. Fig.6 illustrates the test setup. Three stripixel detectors, separated at 50 mm intervals to constitute a hodoscope, were placed inside the dark box that was flushed with flowing dry nitrogen, and the cooling system was set to 0$^{\circ}$ C. The silicon sensor's backplane (ohmic side) of each of the stripixel detectors was supplied with 200 Volts operation voltage via the RCC boards (each RCC contains one FPGA chip. The RCC FPGA board decodes the signals and distributes them to the SVX4 chips on the corresponding ROC). The latter were connected by cables to the Front-End-Module (FEM) prototype, communicating to the data acquisition (DAQ) via the USB interface (each FEM sends timing and control signals and serial data to each RCC board). The event trigger was set by the coincidence of three scintillators placed upstream and downstream of the box, and a "BEAM ON" signal from the accelerator.

We used the same algorithm to analyze the data from the proton-beam test as we applied to the results of the beta-source and cosmic-ray tests. Fig. 7a shows the profile of the proton beam at 120 GeV impinging on the front stripixel detector (layer-1). The beam was perpendicular to the silicon sensor plane. The charge-sharing property in the X-stripixel and U-stripixel is characterized by the asymmetry,

$$A_Q = ( Q_x - Q_u ) / ( Q_x + Q_u ),$$

where $Q_x$ and $Q_u$ represent the collected charges in the X and U stripixels, respectively. For optimum two-dimensional position sensitivity, we expect that, on average, $A_Q$ is zero with a narrow width. The results of the $A_Q$ measurements in the front layer (stripixel detector in front of the incoming proton beam) are plotted in Fig. 7b. They show that the $A_Q$ distribution peaks at

-0.015 with a width of about 0.15, demonstrating the validity of the principle of attaining two-dimensional position sensitivity by charge-sharing.

The energy loss of particles in the silicon sensor (corrected ADC) of the first stripxel detector shown in Figs. 8a, and b, respectively, denote the signal distribution before and after applying the event-by-event pedestal subtraction method. Figs. 8c and d allowed us to calculate the S/N ratios; they are 10.3 and 10.1 for the X and U stripxels, respectively. The S/N ratio in this beam test is slightly higher than in the previous one with the beta source because in the former we implemented in the readout the capability to select the best integration timing in the SVX4 readout chips.

Fig. 9 represents the structure of a single event. In Figs. 9a, b, and c, we observe clearly in the X and U stripxels the particles' energy loss (corrected ADC) during the passage of the proton beam through the three layers (three stripxel detectors). In Fig. 9d, we note that the hits, using this S/N, can be well reproduced by linear-fit track. With this tracking information, Fig. 10 reveals that the hits' residual distributions in the X and U stripxels for the middle detector correspond, respectively, to 0.42 x 80 (μm) = 33.6 (μm) and 0.44 x 80 (μm) = 35.2 (μm) from the RMS values (tracks are defined by layers 1 and 3). Correspondingly, the distribution of tracking efficiency in the X and U stripxels are shown in Figs. 11a and b. We determined tracking efficiencies by looking at the energy deposited in the middle detector at the expected channel (we summed the ADC from the expected channel ±2 neighboring channels) which also was influenced by the other two detectors (layers 1 and 3). For the X stripxel, we extracted a tracking efficiency of 99.5 ± 0.2%, and a value of 98.9 ± 0.2% for the U stripxel,

## 4. Summary and status

The stripxel detector (using the prototype readout card, ROC) demonstrated good performance in exposures to electrons from the beta source ($^{90}$Sr), muons from cosmic-rays, and a proton beam at 120 GeV. The results verified that the hits were detected successfully. We noted a good correlation between the X and U stripxels in the number of hits detected, with a signal-to-noise value of 10.4; tracking efficiencies in both stripxels were very good (better than 98.9 ± 0.2 %). Based on this excellent performance, the status of fabricating the silicon stripxel detectors construction has moved into a pre-production phase.

## Acknowledgments

This research was partially supported by US Department of Energy, DE-AC02-98CH10886. The authors would like to thank D. Pinelli, J. Triolo, and R. Beuttenmuller from BNL's Instrumentation Division at BNL and staffs of the FNAL MTEST Facility and Accelerator Division for their technical support.

## References

[1] R. Nouicer e-Print: arXiv:0901.0910 [nucl-ex] (to be published in the European Physical Journal) and references therein.

[2] I. Arsen et al. Nucl. Phys. A 757, (2005) 1; B.B.Back et al. Nucl. Phys. A 757, (2005) 28; J. Adams et al. Nucl. Phys. A 757, (2005) 102; K. Adcox et al. Nulc. Phys. A 757, (2005) 184.

[3] M. Baker et al., *Proposal for a Silicon Vertex Tracker (VTX) for the PHENIX Experiment*, BNL-72204-2004, Physics Dept. BNL (2004).

[4] T. Roser, "RHIC and its upgrade programs", Proceedings of EPAC08, Genoa, Italy, FRXAGM01 (2008) 3723.

[5] K. Adcox et al., Nucl. Instr. and Meth. A 499 (2003) 469.

[6] W. Snoeys, et al., Nucl. Instr. and Meth. A 466 (2001) 366; K. Wyllie, ALICE1LHCB Document, CERN, 2003.

[7] http://www-cdf.lbl.gov/users/mweber/svx4/

[8] Z. Li, Nucl. Instr. and Meth. A 518 (2004) 738.

[9] R. Nouicer et al. Nucl. Instr. and Meth. B 261 (2007) 1067; R. Nouicer et al. POS VERTEX2007:042, 2007. e-Print: arXiv:0801.2947.

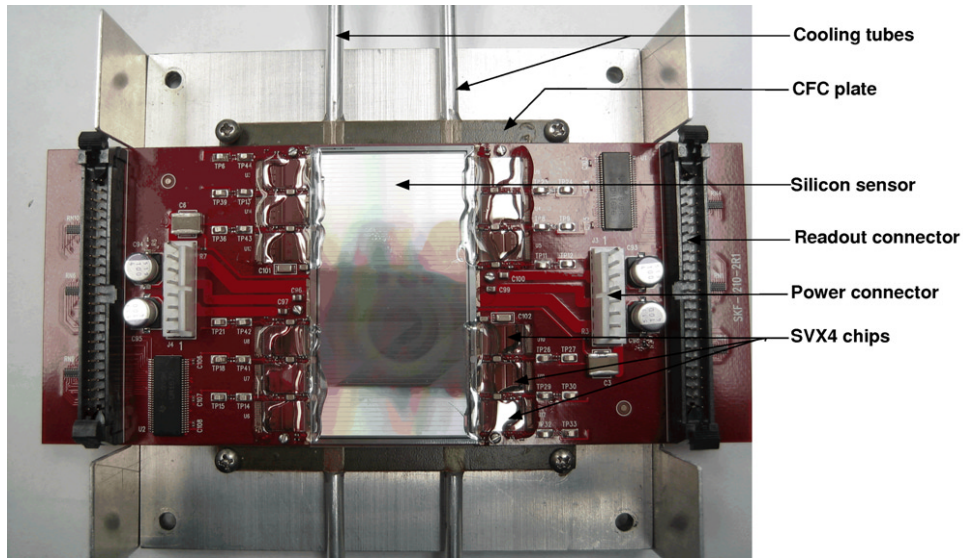

**Fig. 1:** Photograph of the silicon stripixel detector mounted on a thermal plate made from carbon fiber composite (CFC) that contains two cooling tubes.

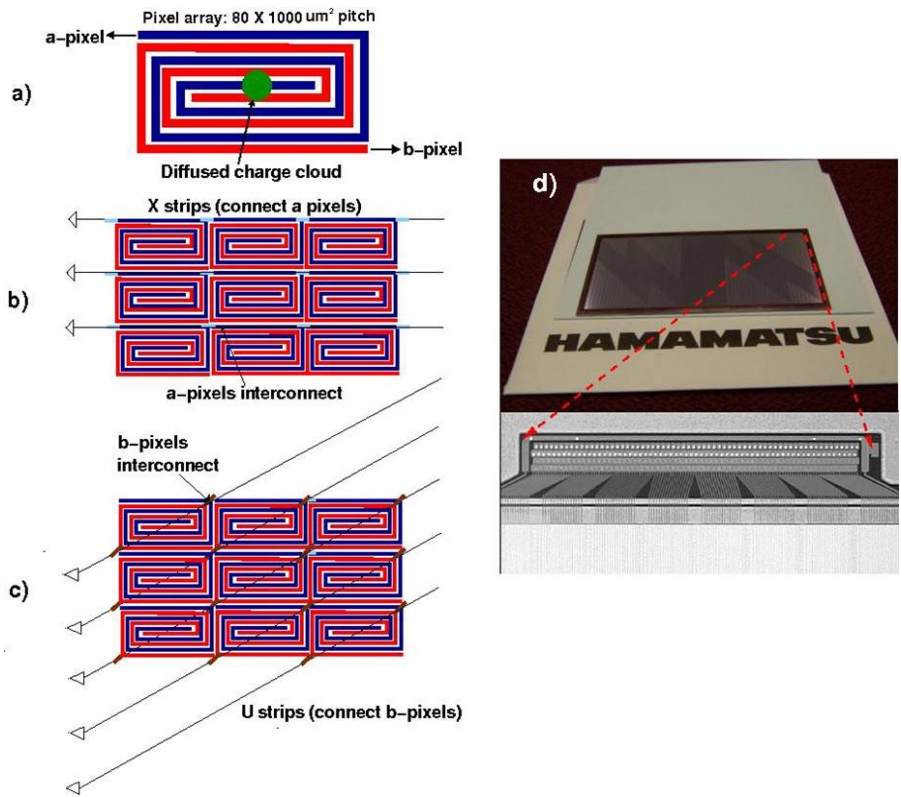

**Fig. 2:** a) Stripixel silicon sensor concept with two spiral-shaped interleaved implants; the a-pixel and b-pixel. b) a-pixels are connected in such way to form a X-stripixel. c) b-pixels are connected to form a U-stripixel. d) shows a photograph of one stripexel silicon sensor (with a magnification of one bonding pad region) fabricated in a commercial full-production process.

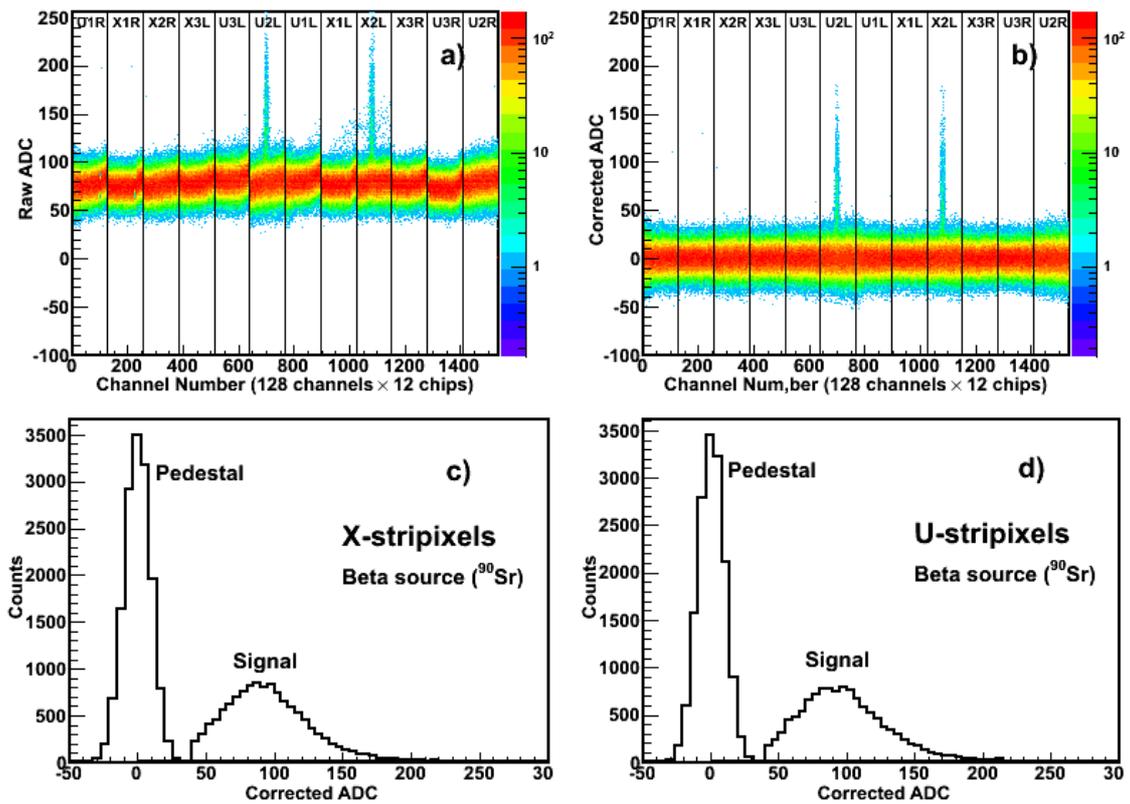

**Fig. 3**: Response of stripixel detector to electrons from beta source ($^{90}$Sr): a) raw ADC versus channel number; and b) corrected ADC versus channel number (see text). c) and d), respectively, show the pedestal and the distribution of electron signals for X and U stripixels.

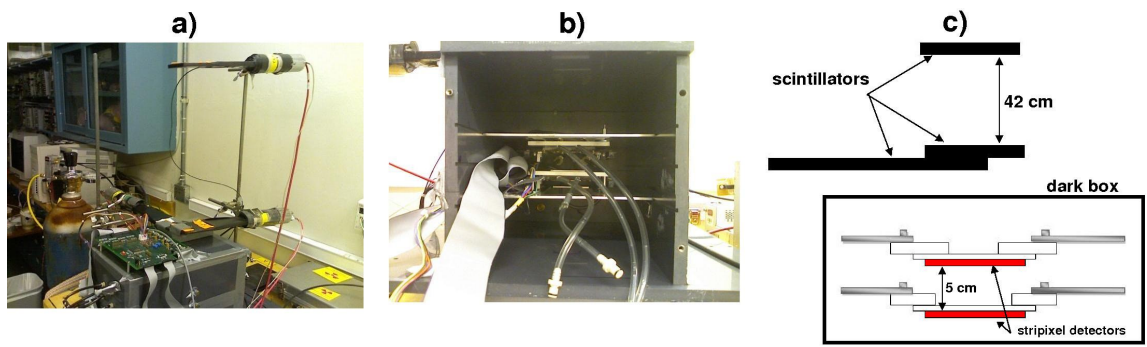

**Fig. 4**: Cosmic-ray test setup. a) outside view of the dark box with scintillators; b) inside view of the dark box (two stripixel detectors), and, c) schematic of the setup.

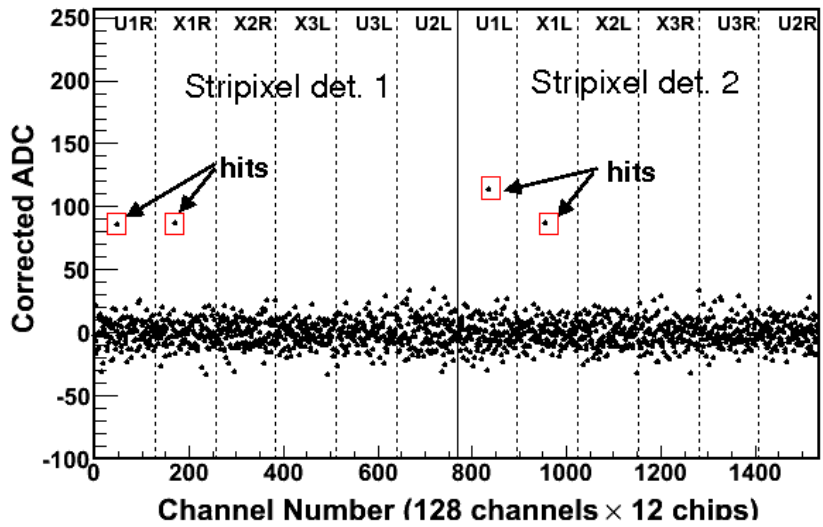

**Fig. 5**: Response of stripixel detectors to one muon from cosmic-ray (a single event). Corrected ADC versus channel number for a single event observed in the hodoscope (two silicon stripixel detectors).

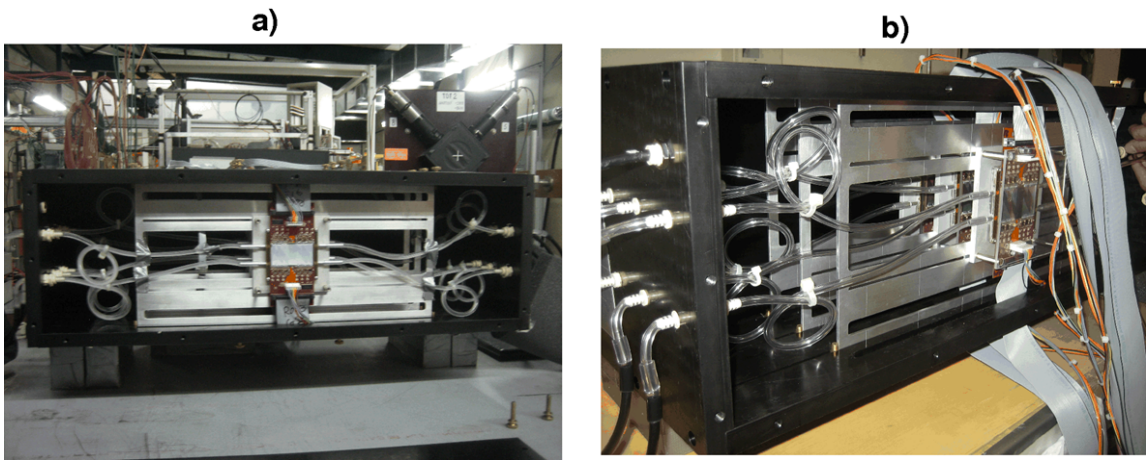

**Fig. 6**: Stripixel detectors setup for proton beam test at the Meson Test Beam (experiment T984) facility at FNAL during August 2008. a) is the front view where the front layer (first stripixel detector) is visible and the incoming 120 GeV proton beam passed first through this layer. b) shows the side view where the three layers (three stripixel detectors) are seen clearly.

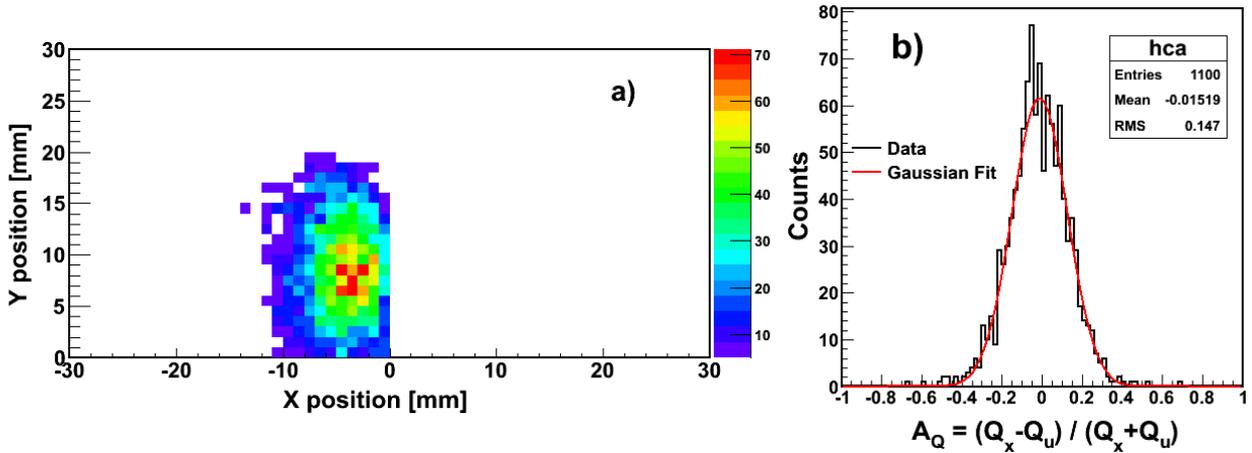

**Fig. 7**: a) profile of the proton beam at 120 GeV on layer 1 (front stripixel detector). b) charged correlations (charge sharing) in between the X- and U-stripixels obtained from the 120 GeV proton beam.

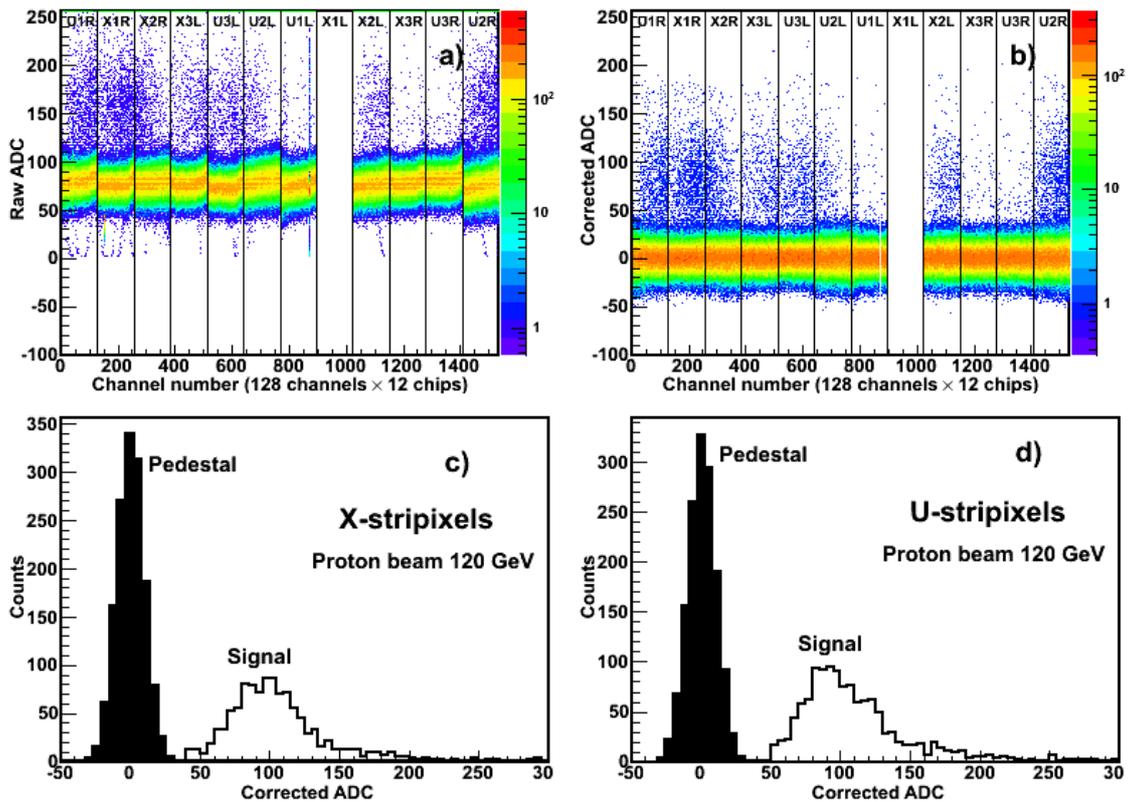

**Fig. 8**: Response of stripixel detector to the proton beam at 120 GeV: a) raw ADC versus channel number; and, b) corrected ADC versus channel number. c) and d) show the pedestal- and proton signal distributions for the X- and U-stripixels, respectively.

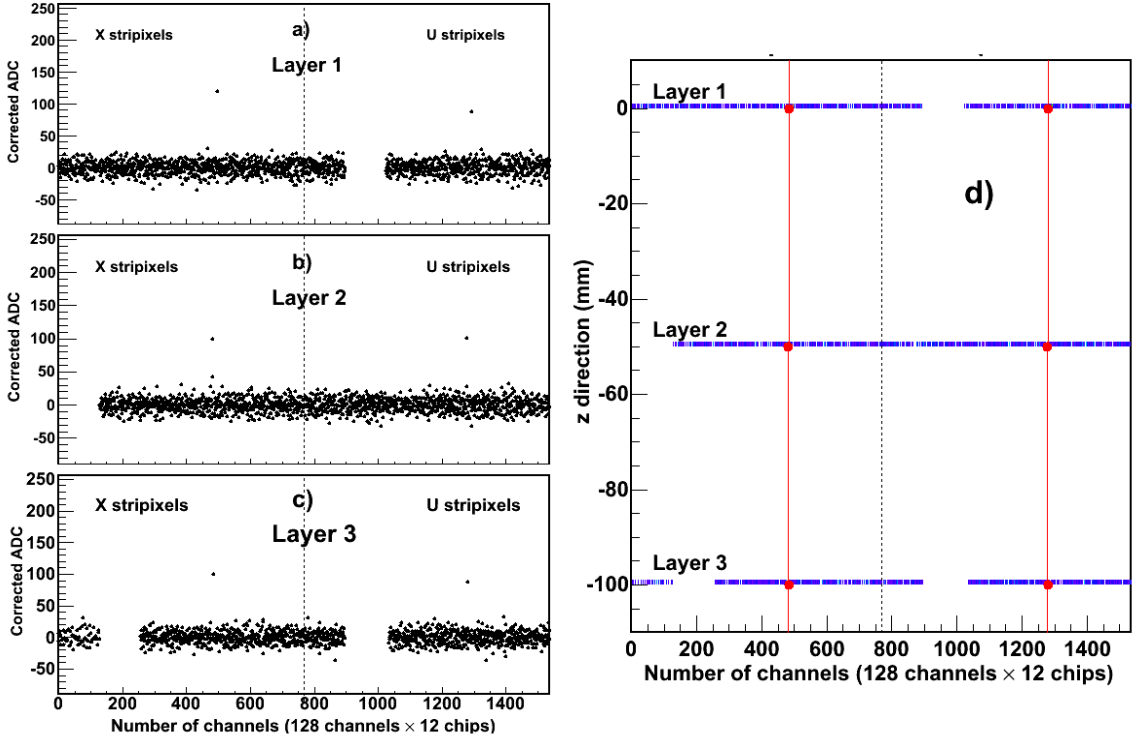

**Fig. 9**: Event structure of a single event from three layers (three stripixel detectors) in proton beam at 120 GeV. a), b), and c) plot the corrected ADC (energy loss) versus channel number for the three layers. Note: The no-hits regions correspond to dead regions (masked the SVX4 chips). d) shows a clustered channel position plotted as circles, points, and tracks by linear fitting, depicted as straight lines for the X- and U- stripixels.

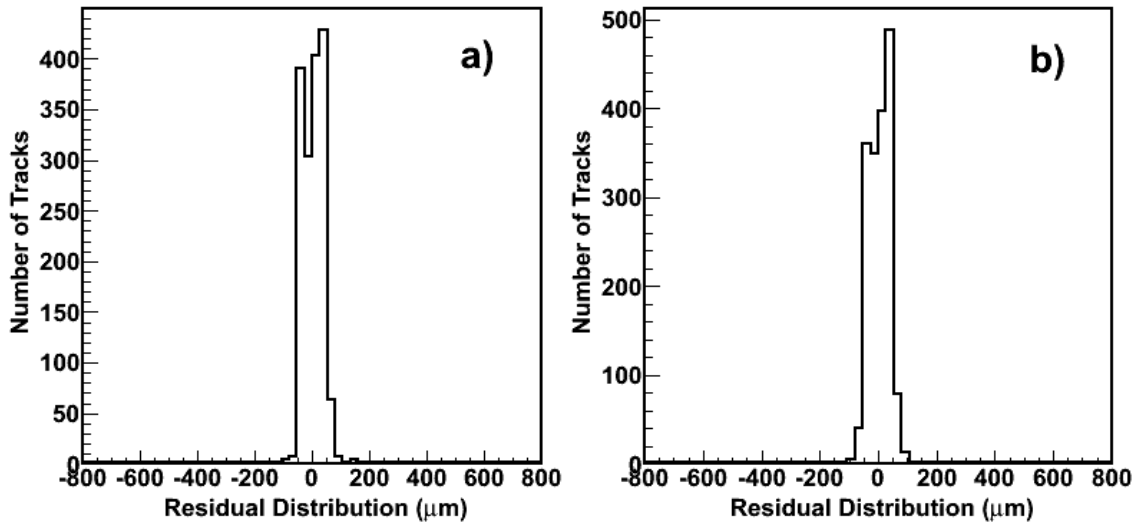

**Fig. 10**: Hit residuals from tracks obtained from the stripixel detectors in the proton beam at 120 GeV. a) and b), respectively, show the hit residuals from the tracks in the X- and U- stripixel detector (middle stripixel detector); they correspond to 0.42 x 80 (μm) = 33.6 (μm), and 0.44 x 80 (μm) = 35.2 (μm) from the RMS values.

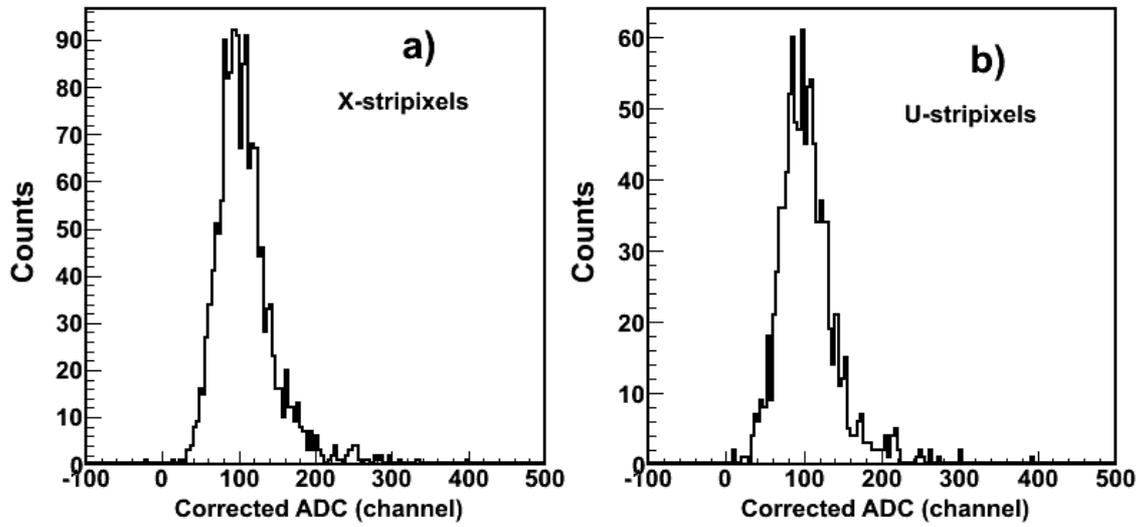

**Fig. 11**: Distributions of tracking efficiency using the stripixel detector (middle layer) in a 120 GeV proton beam. a) and b), respectively, are the clustered ADC distributions for X and U stripixels: the corresponding tracking efficiencies are 99.5 ± 0.2%, and 98.9 ± 0.2%.